\documentclass[11pt,a4paper,english]{amsart}
\usepackage[utf8]{inputenc}
\usepackage[english]{babel} 

\usepackage{amsmath} %Numberwithin
\usepackage{amsthm} %Cambiar lo de teorema, proposición...
\usepackage{graphicx} 
\usepackage{xcolor} %Imprimir palabras de otro color
\usepackage{amsfonts}
\usepackage[biblabel]{cite}
\usepackage{bm} %Letra en negrita en math mode
\usepackage[hidelinks]{hyperref} %Poner links
\usepackage{amssymb} %Usar \supsetneq
\usepackage{tikz-cd} %Diagramas
\usepackage{amscd}
\usepackage{etoolbox}% http://ctan.org/pkg/etoolbox
\patchcmd{\thmhead}{(#3)}{#3}{}{}
\usepackage{enumitem}%quitar indentación enumerate
\usepackage{breqn} %break lines with dmath
\usepackage{faktor} %quotient rings
\usepackage{xfrac} %also quotients
\usepackage{algorithm}
\usepackage{algpseudocode}
\usepackage{blkarray}
\usepackage{pgfplots}
\usepackage{amsaddr}

\everymath=\expandafter{\the\everymath\displaystyle}

\algnewcommand\algorithmicinput{\textbf{Input:}}
\algnewcommand\Input{\item[\algorithmicinput]}
\algnewcommand\algorithmicoutput{\textbf{Output:}}
\algnewcommand\Output{\item[\algorithmicoutput]}

\DeclareMathOperator{\supp}{supp} %Soporte
\DeclareMathOperator{\RM}{RM}

\DeclareMathOperator{\wt}{wt}
\DeclareMathOperator{\rk}{rank}
\DeclareMathOperator{\RS}{RS}
\DeclareMathOperator{\enc}{enc}
\DeclareMathOperator{\Gr}{Gr}
\DeclareMathOperator{\GHW}{GHW}
\DeclareMathOperator{\hierarchy}{hierarchy}

\newcommand{\F}{{\mathbb{F}}}
\newcommand{\fq}{\mathbb{F}_q}

\usepackage{mathtools} %Valor abs y norma. Con * no se ajusta la altura.
\DeclarePairedDelimiter\abs{\lvert}{\rvert}%
\DeclarePairedDelimiter\norm{\lVert}{\rVert}%

\makeatletter
\let\oldabs\abs
\def\abs{\@ifstar{\oldabs}{\oldabs*}}
\let\oldnorm\norm
\def\norm{\@ifstar{\oldnorm}{\oldnorm*}}
\makeatother

\newtheorem{thm}{Theorem}[section]

\newtheorem{cor}[thm]{Corollary}
\newtheorem{lem}[thm]{Lemma}
\theoremstyle{definition}
\newtheorem{defn}[thm]{Definition} 

\newtheorem{rem}[thm]{Remark} 
 
\newtheorem{ex}[thm]{Example}

\title[An algorithm for computing GHWs and the Sage package {\tt GHWs}]{An algorithm for computing generalized Hamming weights and the Sage package {\tt GHWs}}
\author{Rodrigo San-José}
\address{IMUVA-Mathematics Research Institute, Universidad de Valladolid, 47011 Valladolid (Spain)}
\email{rodrigo.san-jose@uva.es}

\thanks{This work was supported in part by the following grants: Grant PID2022-138906NB-C21 funded by MICIU/AEI/10.13039/501100011033 and by ERDF/EU, and FPU20/01311 funded by the Spanish Ministry of Universities.}

\subjclass[2020]{Primary: 94B05. Secondary: 94-04, 11T71}

\keywords{Generalized Hamming weights, Brouwer-Zimmermann algorithm, secret sharing schemes, linear codes}

\begin{document}
\maketitle

\begin{abstract}
We generalize the Brouwer-Zimmermann algorithm, which is the most efficient general algorithm for computing the minimum distance of a random linear code, to the case of generalized Hamming weights. We also adapt this algorithm to compute the relative generalized Hamming weights of a nested pair of linear codes. In the package {\tt GHWs} we provide an implementation of this algorithm in Sage, as well as several other utilities for working with generalized Hamming weights. With this implementation, we show that the proposed algorithm is faster than the naive approach of computing the generalized Hamming weights using the definition. 
\end{abstract}

\section{Introduction}

Linear error correcting codes were introduced to study the problem of sending digital information through a noisy channel \cite{macwilliams}. Linear codes are able to correct errors by adding redundancy to the message, and they have three main basic parameters, namely the length $n$, the dimension $k$, and the minimum distance $d$. The rate $k/n$ determines how much information the code can transmit per bit (the higher the rate, the lower redundancy we require), and $d/n$ determines the proportion of errors the code can correct. These parameters are related (for example, via the Singleton bound $d\leq n-k+1$), and finding codes with an optimal trade-off is a difficult problem in general. Over time, linear codes have found many additional applications. To determine the performance of the code for each of them, additional properties have to be considered. Examples of such applications are locally recoverable codes \cite{LRCOriginal}, code-based cryptography \cite{McEliece1978,niederreiterCryptosystem}, or quantum codes \cite{calderbankgoodquantum,cssoriginal2}, in which one has to study locality parameters and orthogonality properties.

The generalized Hamming weights (GHWs) of a linear code, first defined in \cite{weiGHW}, generalize the notion of minimum distance. In the original paper \cite{weiGHW}, it is shown that these parameters characterize the performance of the code on the wire-tap channel of type II and as a $t$-resilient function. In \cite{guruswammiGHWlistdecoding,guruswammiGHWlistdecodingTensorInterleaved}, the connection with list decoding is established, and it is used to derive better bounds for the list size. Related to the GHWs, the relative generalized Hamming weights (RGHWs) of a nested pair of codes were introduced as an extension of GHWs, mainly to characterize the security of linear ramp secret-sharing schemes \cite{luoPropertiesRGHWs,matsumotoRGHW}. It is widely known that the first RGHW, also known as relative minimum distance, determines the minimum distance of quantum codes \cite{kkks}. However, the second RGHW also determines the minimum distance of the quantum codes obtained via Steane enlargement \cite{steaneSteaneEnlargement}, as seen in \cite[Lem. 5]{hamadaSteaneEnlargement}. 

These applications have motivated the study of codes with algebraic structure that allows to derive properties about their GHWs and their RGHWs. Some classical results concern cyclic codes \cite{weiGHWcyclic,janwaGHWcyclic,zhangRGHWcyclic,yangGHWcyclic,xiongGHWcyclic} and Reed-Muller codes \cite{pellikaanGHWRM,geilRGHWReedMuller}, as well as algebraic geometry codes \cite{munueraGHWhermitica,stichtenothGHWAG,geilRGHWAGcodes}, and there have been many recent advances with some well-known classes of codes, such as projective Reed-Muller codes \cite{beelenGHWPRM,sanjoseRecursivePRM}, Cartesian codes \cite{beelenGHWcartesian,dattaRGHWcartesian}, hyperbolic codes \cite{eduardoGHWHyperbolic}, matrix-product codes \cite{sanjoseGHWMPC} or norm-trace codes \cite{sanjoseGHWNT}.

However, if we consider a random linear code, that is, without any known structure, the computation of its minimum distance is an NP hard problem \cite{vardyIntractability}. The fastest general algorithm for computing the minimum distance of a random code is the Brouwer-Zimmermann algorithm \cite{zimmermann} (also see \cite{grasslBrouwerZimmermann} for a review of this algorithm and improvements). Since the GHWs generalize the minimum distance, and the RGHWs extend the GHWs, their computation is at least as intractable as that of the minimum distance. To the best knowledge of the author, there is currently no efficient implementation or algorithm available in the literature for computing the GHWs of a random linear code over an arbitrary finite field. 

The goal of this paper is to generalize the Brouwer-Zimmermann algorithm for computing the GHWs of a random linear code and the RGHWs of a nested pair of linear codes, and give a general implementation of this algorithm. We provide this implementation in the form of a Sage package \cite{sagemath,githubGHWs}, which also includes additional functions related to GHWs, such as a function to compute the higher weight spectra of a linear code. The aim of this package is to provide a general implementation that works for any linear code. In particular, this means that the implementation should work for any finite field, and therefore we have to use specialized mathematical software. The most common ones for working with linear codes are Magma \cite{magma}, GAP \cite{gap,guava} and Sage \cite{sagemath}. We decided to use Sage because it is open-source and it is based on Python, making it easier to analyze and understand, which is another important objective of this package, since we would like this work to pave the way for research on how to compute the GHWs of a linear code more efficiently.

The structure of the paper is as follows. In Section \ref{s:preliminaries} we give the necessary mathematical background related to linear codes and their GHWs and RGHWs. Section \ref{s:BZalg} explains how to generalize the Brouwer-Zimmermann algorithm to this setting. In Section \ref{s:enumeration}, we study how to construct subspaces of a given dimension with increasing size of their support. This problems translates to constructing all possible matrices in reduced row echelon form with a certain ``shape'', and we show how to implement the construction of these matrices. Section \ref{s:implementation} describes the Sage implementation of our algorithms. We explain which functions are included in the package and we give more details on how exactly we have implemented certain aspects of the algorithm. We test our implementation in Section \ref{s:tests}, where we first explain how we can check that the results of the implementation are correct, and then we conduct a performance analysis. Since currently there are no other alternatives, we compare our implementation with a naive implementation, using directly the definition of the GHWs. Our tests show that the difference in performance is considerable, e.g., we need to use a logarithmic scale for the time in the graphics since the difference in speed is exponential. We include in the package an alternative implementation of some of the functions which requires less RAM usage, and we study the difference in performance with respect to the standard implementation. We also compare the time required to compute the GHWs independently one by one (using the function {\tt GHW}), and computing them using the information of the previous ones (using the function {\tt hierarchy}). Finally, in Section \ref{s:conclusion}, we summarize our results and propose new avenues of research.

\section{Preliminaries}\label{s:preliminaries}
Let $q=p^s$ be a power of a prime $p$. We denote by $\fq$ the finite field with $q$ elements. We say that $C$ is a linear code of length $n$ and dimension $k$ if it is a $k$ dimensional linear subspace of $\fq^n$. A matrix $G\in \fq^{k\times n }$ is a generator matrix of $C$ if its rows form a basis of $C$. The dual code of $C$ is denoted by $C^\perp$, and is the orthogonal space of $C$ with respect to the usual Euclidean inner product. Thus, we have $\dim C^\perp=n-k$. A generator matrix $H\in \fq^{(n-k)\times n }$ of $C^\perp$ is called a parity check matrix for $C$. The main parameters of the code $C$ are the length $n$, the dimension $k$, and the minimum distance $d$, and we say that $C$ is an $[n,k,d]$ code. The parameter $d$ is defined in terms of the Hamming weight. Given $c\in \fq^n$, we define its Hamming weight as
$$
\wt(c):=\abs{\{i : c_i\neq 0\}}.
$$
The minimum distance $d=\wt(C)$ of $C\subset \fq^n$ is then defined as
$$
\wt(C):=\min \{\wt(c):c\in C\}. 
$$
The naive method to compute $\wt(C)$ would be to enumerate all the codewords $c\in C$, and then compute the minimum of their weights. Since $\wt(c)=\wt( \lambda c)$ for any $\lambda \in \fq\setminus \{0\}$, we would need to enumerate 
$$
\frac{q^k-1}{q-1}
$$
codewords (this is the number of nonzero vectors of $\fq^k$ up to multiples). Now we define the GHWs of a linear code. For this, we have to define first the support of a linear subspace. Given $D\subset \fq^n$, the support of $D$ is
$$
\supp(D):=\left\{ 1\leq i \leq n: \exists c \in D \textnormal{ with } c_i \neq 0 \right\}.
$$

Note that if $G_D$ is a generator matrix for $D$, that is, a matrix whose rows form a basis for $D$, and $\{g_1,\dots,g_r\}$ are the rows of $G_D$, then  
\begin{equation}\label{eq:supp}
\supp(D)=\bigcup_{i=1}^r \supp(g_i).
\end{equation}
In particular, this shows that computing the support of a subcode $D$ from a generator matrix $G_D$ is straightforward.

\begin{defn} \label{def:ghw}
The $r$th generalized Hamming weight of an $[n,k,d]$ code $C$ is 
$$
d_r(C):=\min \left\{ \abs{ \supp(D) }: D \textnormal{ is a subcode of } C \textnormal{ of dimension } r \right\},
$$
where $1\leq r \leq k$. The weight hierarchy of $C$ is the set 
$$
\left\{ d_r(C): 1\leq r \leq k \right\}. 
$$
\end{defn}

\begin{rem}
If $r=1$, the subspaces $D$ of $C$ with dimension 1 are precisely the codewords (up to multiples). The support of a codeword is given by the positions of its nonzero coordinates, and its cardinality is precisely the number of nonzero coordinates, that is, its Hamming weight. Thus, we have $d_1(C)=\wt(C)$. 
\end{rem}

From \cite{weiGHW} we have the following general properties of the GHWs of a code.

\begin{thm}[(Monotonicity)]\label{t:monotonia}
For an $[n,k]$ linear code $C$ with $k>0$ we have
$$
1\leq d_1(C)<d_2(C)<\cdots <d_k(C)\leq n.
$$
\end{thm}
\begin{cor}[(Generalized Singleton Bound)]\label{c:singletongeneralizada}
For an $[n,k]$ linear code $C$ we have
$$
d_r(C)\leq n-k+r, \; 1\leq r\leq k.
$$
\end{cor}

\begin{thm}[(Duality)]\label{t:ghwdual}
Let $C$ be an $[n,k]$ code. Then
$$
\{d_r(C):1\leq r\leq k\}=\{1,2,\dots,n\}\setminus \{n+1-d_r(C^\perp):1\leq r\leq n-k\}.
$$
\end{thm}

\begin{rem}\label{r:GHWMDS}
As a consequence of Theorem \ref{t:monotonia} and Corollary \ref{c:singletongeneralizada}, if $C$ is an MDS code with parameters $[n,k,d]$ (that is, we have $d=n-k+1$), then we have
$$
d_r(C)=n-k+r
$$
for any $1\leq r \leq k$. 
\end{rem}

The RGHWs, which extend the GHWs, were introduced in \cite{luoPropertiesRGHWs,matsumotoRGHW}. This was in the context of linear ramp secret-sharing schemes, which can be constructed from a pair of nested linear codes. In these schemes, a dealer distributes shares of a secret to a certain number of participants. Using the shares of a sufficiently large number of participants, it is possible to recover the secret. A ramp secret sharing scheme is said to have $(t_1,\dots,t_\ell)$-privacy and $(h_1,\dots,h_\ell)$-reconstruction if $t_1,\dots,t_\ell$ are the largest possible, and $h_1,\dots,h_\ell$ the smallest possible, such that:
\begin{itemize}
    \item no set of $t_m$ participants can obtain $m$ $q$-bits of information about the secret,
    \item any set of $h_m$ participants can obtain $m$ $q$-bits of information about the secret. 
\end{itemize}
These parameters can be completely characterized in terms of the RGHWs of the pair of linear codes used for the secret sharing scheme, as well as the RGHWs of their duals \cite{geilRGHWAGcodes}. Although this is the application in which the RGHWs are used with complete generality, for $r=1$ (relative minimum distance), they determine the error-correction capability of quantum codes \cite{kkks}. Moreover, for $r=2$, they also determine the minimum distance of the codes derived via Steane enlargement \cite{steaneSteaneEnlargement,hamadaSteaneEnlargement}. We introduce now their definition.

\begin{defn} \label{d:rghw}
Let $C_2\subset C_1 \subset \fq^n$ be two linear codes, and $k_1=\dim C_1$, $k_2=\dim C_2$. Let $r$ with $1\leq r \leq k_1-k_2$. The $r$th relative generalized Hamming weight of $C_1$ and $C_2$, denoted by $M_r(C_1,C_2)$, is
$$
M_r(C_1,C_2)=\min \{ \abs{\supp(D)} : D \text{ is a subcode of $C_1$ with } \dim D=r, \; D\cap C_2=\{0\}\}.
$$
The relative weight hierarchy of $C_1$ with respect to $C_2$ is the set
$$
\{ M_r(C_1,C_2),\; 1\leq r\leq k_1-k_2\}.
$$
\end{defn}

It is clear from the definition that the RGHWs are a generalization of the GHWs obtained by considering $C_2=\{0\}$. Note that, for $1\leq r \leq k_1-k_2$, we have
$$
M_r(C_1,C_2)\geq d_r(C_1).
$$

The RGHWs satisfy similar properties to GHWs, as shown in \cite{luoPropertiesRGHWs}.

\begin{thm}\label{t:monotoniarghw}
Let $C_2\subset C_1\subset \fq^n$ be linear codes with $\dim C_1=k_1$, $\dim C_2=k_2$. Then
$$
1\leq M_1(C_1,C_2)< M_2(C_1,C_2)<\cdots < M_{k_1-k_2}(C_1,C_2)\leq n.
$$
Moreover, for any $1\leq r \leq k_1-k_2$, we have
$$
M_r(C_1,C_2)\leq n-k_1+r.
$$
\end{thm}

\begin{rem}\label{r:lowerboundGHW}
The monotonicity from Theorem \ref{t:monotonia} (resp. Theorem \ref{t:monotoniarghw}) implies $d_r(C)\geq r$ (resp. $M_r(C_1,C_2)\geq r$). 
\end{rem}

One may wonder if there is a duality result similar to Theorem \ref{t:ghwdual} for RGHWs. This is one example in which having a package such as {\tt GHWs} can be useful for research, since we can look for random codes and see if it is possible to have such a result. In the next example we show two pairs of codes with the same relative weight hierarchy, but different relative weight hierarchy for the duals, which implies that, in general, the RGHWs of a pair of codes do not determine the RGHWs of their duals.

\begin{ex}
Let $q=2$ and consider the codes $C_1,C_2,C'_1,C'_2$ generated by the matrices
$$
G_1=\left(\begin{array}{rrrrrrrrrr}
0 & 1 & 0 & 1 & 0 & 0 & 1 & 0 & 0 & 0 \\
1 & 1 & 1 & 1 & 1 & 1 & 1 & 0 & 1 & 0 \\
0 & 0 & 0 & 0 & 0 & 0 & 1 & 1 & 0 & 1 \\
1 & 0 & 0 & 1 & 0 & 0 & 0 & 0 & 0 & 0 \\
0 & 0 & 1 & 1 & 0 & 1 & 0 & 0 & 0 & 0
\end{array}\right),\; G_2=\left(\begin{array}{rrrrrrrrrr}
0 & 1 & 0 & 1 & 0 & 0 & 1 & 0 & 0 & 0 \\
1 & 1 & 1 & 1 & 1 & 1 & 1 & 0 & 1 & 0 \\
0 & 0 & 0 & 0 & 0 & 0 & 1 & 1 & 0 & 1
\end{array}\right),
$$
$$
G'_1=\left(\begin{array}{rrrrrrrrrr}
1 & 1 & 0 & 1 & 0 & 0 & 0 & 0 & 0 & 1 \\
0 & 1 & 0 & 1 & 1 & 1 & 0 & 1 & 0 & 0 \\
1 & 0 & 1 & 0 & 0 & 0 & 1 & 0 & 1 & 0 \\
1 & 1 & 1 & 0 & 0 & 0 & 0 & 0 & 0 & 0 \\
0 & 1 & 0 & 1 & 0 & 0 & 0 & 0 & 0 & 0
\end{array}\right),\; G'_2=\left(\begin{array}{rrrrrrrrrr}
1 & 1 & 0 & 1 & 0 & 0 & 0 & 0 & 0 & 1 \\
0 & 1 & 0 & 1 & 1 & 1 & 0 & 1 & 0 & 0 \\
1 & 0 & 1 & 0 & 0 & 0 & 1 & 0 & 1 & 0
\end{array}\right),
$$
respectively. It is straightforward to check that $C_2\subset C'_1$ and $C_2'\subset C_1$. We have $M_1(C_1,C_2)=M_1(C'_1,C'_2)=2$ and $M_2(C_1,C_2)=M_2(C'_1,C'_2)=4$. In other words, the pairs $(C_1,C_2)$ and $(C'_1,C'_2)$ have the same relative weight hierarchy. If we consider now the pairs $(C_2^\perp,C_1^\perp)$ and $((C'_2)^\perp,(C'_1)^\perp)$, we have $M_1(C_2^\perp,C_1^\perp)=M_1((C'_2)^\perp,(C'_1)^\perp)=2$, but $M_2(C_2^\perp,C_1^\perp)=3\neq 4= M_2((C'_2)^\perp,(C'_1)^\perp)$. That is, the dual pairs do not have the same relative weight hierarchy. 
\end{ex}

Another object that has been studied in this context is the higher weight spectrum \cite{ghorpadeHigherWeightRM,kaplanHigherWeightCurves,johnsenHigherWeightVeronese,kaipaHigherWeightVeronese3fold,johnsenHigherWeightVeroneseThreefolds,doughertyHigherWeights}, a generalization of the weight spectrum. For $1\leq r \leq k$ and for each $w=d_r(C),\dots,n$, we denote by $A_w^{(r)}(C)$ the number of subcodes $D\subset C$ with $\dim D=r$ and $\abs{\supp(D)}=w$. The ordered multiset $\{A_w^{(r)}(C),\; d_r(C)\leq w\leq n\}$ is the $r$th weight spectrum of $C$. The multiset of $r$th weight spectra for $r=0,\dots,k$ is the higher weight spectra of $C$. Note that for this, we consider $d_0(C)=0$ and $A_0^{(0)}(C)=1$. We can similarly define the $r$th relative weight spectrum and higher weight spectra of $C_1$ with respect to $C_2$ by only considering subspaces $D\subset C$ such that $D\cap C_2=\{0\}$. 

\section{Brouwer-Zimmermann-like algorithm} \label{s:BZalg}
We consider a code $C\subset \fq^n$ with $k=\dim C$, and a generator matrix $G$ of $C$. Given a subspace $E\subset \fq^k$, we denote by $\enc_G(E)\subset C$ the subspace obtained as the image of $E$ by the linear transformation defined by $G$. Let $\Gr(r,\fq^k)$ be the set of all subspaces of $\fq^k$ with dimension $r$ (the notation comes from the Grassmannian). A naive algorithm to compute the GHWs of C would consist in enumerating all the elements of $\Gr(r,\fq^k)$, and then computing the minimum of the supports of $\enc_G(E)$, for every $E\in \Gr(r,\fq^k)$. The number of such subspaces is given by the following Gaussian binomial:
$$
{ k\brack r}_q=\frac{(q^k-1)(q^k-q)\cdots (q^k-q^{r-1})}{(q^r-1)(q^r-q)\cdots (q^r-q^{r-1})}.
$$
Since this number grows considerably when increasing $q$ and $k$, any method that allows us to not have to check all of these subspaces can provide a significant computational advantage. The Brouwer-Zimmermann algorithm for the minimum distance considers the codewords with increasing weight, i.e., it first considers the vectors of $\fq^k$ with weight 1, then those of weight 2, etc. The equivalent procedure in this case would correspond to considering subspaces $E\subset \fq^k$ with increasing cardinality of the support, i.e., first we consider the subspaces $E$ with $\abs{\supp(E)}=r$ (recall Remark \ref{r:lowerboundGHW}), then subspaces $E$ with $\abs{\supp(E)}=r+1$, etc. Let $w\geq r$, and consider
$$
E_w^r:=\{ E\subset \fq^k, \; \dim E=r, \; \abs{\supp(E)}=w \}.
$$
We have the disjoint union
$$
\Gr(r,\fq^k)=\bigsqcup_{w=r}^k E_w^r.
$$
We explain now the Brouwer-Zimmermann-like algorithm, similarly to the way the Brouwer-Zimmermann algorithm is introduced in \cite{grasslBrouwerZimmermann}. The proof of the following result is straightforward from the definitions (also recall (\ref{eq:supp})).

\begin{lem}\label{l:GIA}
Let $G=(I_k,A)$ a generator matrix for a code $C\subset \fq^n$ and $E\subset \fq^k$ with $\dim E=r$. Let $G_E$ be a generator matrix for $E$. Then 
$$
G_D:=G_EG=(G_E,G_EA)
$$
is a generator matrix for a subcode $D\subset C$ with $\dim D=r$ and $\abs{\supp(D)}\geq \abs{\supp(E)}$.
\end{lem}

\begin{rem}
The previous result can also be generalized to other contexts, for example for rank-metric codes (or sum-rank metric codes). However, in those contexts, it seems that the improvements that will be explained later in this paper by considering several generator matrices cannot be implemented.
\end{rem}

Assume we have computed 
$$
d'=\min_{r\leq i \leq w} \{\abs{\supp(\enc_G(E))}:E\in E_i^r\}
$$
for some $w <k$. Then $d\leq d'$ and, by Lemma \ref{l:GIA}, for every $E\in E^r_\ell$, with $w<\ell$, we must have
$$
\abs{\supp(E)}\geq \ell \geq w+1.
$$
If $w+1\geq d'$, we obtain $d_r(C)=d'$. Therefore, by enumerating subspaces with increasing support, we have a lower bound for the cardinality of the support of the remaining subspaces. This lower bound can be improved if, instead of only considering one generator matrix, we consider more. First, we introduce the notation of information sets. Let $G$ be a generator matrix for $C$. We say that $I\subset \{1,\dots,n\}$ is an information set if the set of columns of $G$ given by the indices of $I$ is linearly independent. In that case, we may always find a generator matrix $G'$ of $C$ such that if we consider only the columns corresponding to $I$, we obtain the identity matrix. In other words, after a permutation, we have $G'=(I_k,A)$. The Brouwer-Zimmermann takes advantage of using several matrices $G_j$ of that form ($G_j$ is a generator matrix for $C$ such that the columns corresponding to $I_j$ give $I_k$), for disjoint information sets $I_1,\dots, I_m$. The argument is as follows. If we have computed 
$$
d'=\min_{r\leq i \leq w} \left\{ \min_{1\leq j \leq m}  \{\abs{\supp(\enc_{G_j}(E))}:E\in E_i^r\} \right\},
$$
for some $w<k$, then for every $E\in E^r_\ell$, with $w<\ell$, such that $\enc_{G_j}(E)$ has not been enumerated before, we have
$$
\abs{\supp(\enc_{G_j}(E))}\geq m(w+1),
$$
for any $j=1,\dots,m$. Indeed, fix $j$ and consider $\supp(\enc_{G_j}(E)))\cap I_i$, for some $1\leq i \leq m$, and assume the subspace $\enc_{G_j}(E)$ has not been enumerated yet. The cardinality of this intersection is greater than or equal to $w+1$ (otherwise it would have been already enumerated as $\enc_{G_i}(E')$, for some $E'\in E^r_{\ell'}$ with $\ell'\leq w$). Since this is true for every $i=1,\dots,m$, and the sets $I_1,\dots, I_m$ are disjoint, we conclude the inequality given above. 

A way to compute these information sets is to start with a generator matrix of the form $G_1=(I_k,A)$ (this can always assumed by permuting), which gives $I_1=\{1,\dots,k\}$, and then using Gaussian elimination to obtain $A=(I_k,A')$ (up to permutation), resulting in a generator matrix $G_2=(B,I_k,A')$ (up to permutation), etc. At some point, we may encounter, for example, that $\rk(A)<k$. In this case, we have to reuse some columns to obtain an information set. In this way we obtain $m$ information sets $I_1,\dots, I_m$, such that $\bigcup_{j=1}^m I_j=\{1,\dots,n\}$, but they are not necessarily disjoint (one can check \cite{grasslBrouwerZimmermann} for the details). For each $1\leq i \leq m$, we define the redundancy of $I_i$ as 
$$
R_i=\abs{I_j\cap \bigcup_{t=1}^{j-1}I_t}.
$$
Arguing as we did with the case of disjoint information sets, when using non-disjoint information sets, each generator matrix $G_i$ contributes $(w+1)-R_i$ to the lower bound. This leads to Algorithm \ref{alg:ghw}, which shows the basic idea of all the algorithms that we will consider. Note that the cost of computing the information sets, and their corresponding generator matrices, is negligible compared to the rest of the algorithm, and we may omit it from the pseudo-code.

\begin{algorithm}
\caption{Brouwer-Zimmermann-like algorithm for the $r$th GHW}\label{alg:ghw}
\begin{algorithmic}[1] %[1] for numbering of lines
\Input $C\subset \fq^n$, $1\leq r \leq k$.
\Output $d_r(C)$.
\State $\GHW_{l} = r$
\State $\GHW_{u} = n-k+r$
\While{$w\leq k$ and $\GHW_{l}<\GHW_{u}$}
    \For{$E\in E^r_w$}
        \For{$j=1,\dots,m$}
            \State $\GHW_{u} = \min \{\GHW_{u}, \abs{\supp(\enc_{G_j}(E))}\}$
        \EndFor
    \EndFor
    \State $\GHW_l=\sum_{j=1}^m \max\{0,w+1-R_j\}$
    \State $w=w+1$
\EndWhile
\State \Return $\GHW_u$
\end{algorithmic}
\end{algorithm}

There is one aspect of Algorithm \ref{alg:ghw} (and Algorithm \ref{alg:hierarchy}) that has not been covered yet, which is efficiently constructing $E_w^r$. We study this problem in Section \ref{s:enumeration}, where we show how to obtain all the subspaces of $E_w^r$ without repeating them.

With respect to RGHWs, we can also use Algorithm \ref{alg:ghw}, except that whenever we find $E\subset \fq^k$ such that $\abs{\supp(\enc_{G_j}(E))}<\GHW_{u}$, we first check if $\enc_{G_j}(E))\cap C_2=\{0\}$, and if that is the case, we set $\GHW_u=\abs{\supp(\enc_{G_j}(E))}$; if that is not the case, we continue with the algorithm, without updating the value of $\GHW_u$. 

Regarding the weight hierarchy, one could just use Algorithm \ref{alg:ghw} for each value of $1\leq r \leq \dim C$. However, we can improve the performance of the algorithm by considering the monotonicity of the GHWs from Theorem \ref{t:ghwdual} (resp. Theorem \ref{t:monotoniarghw} for RGHWs), since it gives that $d_{r-1}(C)+1$ (resp. $M_{r-1}(C_1,C_2)$+1) is a lower bound for $d_r(C)$ (resp. $M_r(C_1,C_2)$). Algorithm \ref{alg:hierarchy} gives the corresponding algorithm to compute the weight hierarchy. It can be modified to compute the relative weight hierarchy in a similar way as one can modify Algorithm \ref{alg:ghw} to compute RGHWs. 

\begin{algorithm}
\caption{Brouwer-Zimmermann-like algorithm for the weight hierarchy}\label{alg:hierarchy}
\begin{algorithmic}[1] %[1] for numbering of lines
\Input $C\subset \fq^n$.
\Output $\{ d_r(C),\; 1\leq r \leq k=\dim C\}$.
\State $\hierarchy = [\;] $
\For {$r=1,\dots,k$}
    \If {$r=1$}
        \State $\GHW_{l} = 1$
    \Else
        \State $\GHW_{l} = \hierarchy[r-1]+1$
    \EndIf
    \State $\GHW_{u} = n-k+r$
    \While{$w\leq k$ and $\GHW_{l}<\GHW_{u}$}
        \For{$E\in E^r_w$}
            \For{$j=1,\dots,m$}
                \State $\GHW_{u} = \min \{\GHW_{u}, \abs{\supp(\enc_{G_j}(E))}\}$
            \EndFor
        \EndFor
        \State $\GHW_l=\sum_{j=1}^m \max\{0,w+1-R_j\}$
        \State $w=w+1$
    \EndWhile
    \State Append $\GHW_{u}$ to hierarchy
\EndFor
\State \Return hierarchy
\end{algorithmic}
\end{algorithm}

\section{Enumerating and constructing subspaces with a given support}\label{s:enumeration}

For both Algorithms \ref{alg:ghw} and \ref{alg:hierarchy} we need to construct the sets $E^r_w$, for $r\leq w\leq k$. If we denote $e_w^r=\abs{E_w^r}$, then we have
$$
e_w^r=\binom{k}{w}\left( { w\brack r}_q-\binom{w}{1}{ w-1\brack r}_q+ \binom{w}{2}{ w-2\brack r}_q +\cdots + (-1)^{w-r} \binom{w}{w-r}{ r\brack r}_q\right).
$$
This is obtained by first considering a particular support of cardinality $w$ (there are $\binom{k}{w}$), then considering all subspaces in $\fq^w$ with dimension $r$ (there are ${ w\brack r}_q$), and finally discarding those subspaces without full support. This can be done by means of the inclusion-exclusion principle: first we discard those subspaces contained in the hyperplane $\{x_i=0\}$, for some $i=1,\dots,w$ (there are $\binom{w}{1}$ such hyperplanes, and for each hyperplane we have ${ w-1\brack r}_q$ subspaces contained in it), which means we have removed the ones contained in two of those hyperplanes twice, etc. By the definitions, we have the relation
$$
{ k\brack r}_q=\sum_{w=r}^k e_w^r.
$$

\begin{rem}
For the case of the minimum distance ($r=1$), note that
$$
{ \ell\brack 1}_q=\frac{q^\ell-1}{q-1}=q^{\ell -1}+q^{\ell-2}+\cdots +1,
$$
and
$$
\begin{aligned}
e^1_w=&\binom{k}{w}\left( \frac{q^w-1}{q-1}-\binom{w}{1}\frac{q^{w-1}-1}{q-1}+\cdots +(-1)^{w-1}\binom{w}{w-1}\frac{q-1}{q-1}\right)\\
=&\frac{\binom{k}{w}}{q-1}\left( q^w-\binom{w}{1}q^{w-1}+\binom{w}{2}q^{w-2}+\cdots+(-1)^{w-1}\binom{w}{w-1}q \right.\\
&-1+\left. \binom{w}{1}-\binom{w}{2}+\cdots +(-1)^{w}\binom{w}{w-1} \right)   \\
=&\frac{\binom{k}{w}}{q-1}\left( (q-1)^w -(-1)^w-(-1)^{w+1} \right)=\binom{k}{w}(q-1)^{w-1},
\end{aligned}
$$
where we have used that $(1-1)^w=0=\sum_{i=0}^w \binom{w}{i}(-1)^i$. It is clear that this is precisely the number of codewords with hamming weight $w$ up to multiples, and we have
$$
\sum_{w=1}^k e^1_w=\sum_{w=1}^k \binom{k}{w}(q-1)^{w-1}=\frac{1}{q-1}\left( \sum_{w=1}^k \binom{k}{w}(q-1)^{w}\right)=\frac{((q-1)+1)^k-1}{q-1}=\frac{q^k-1}{q-1},
$$
which is the number of nonzero vectors in $\fq^k$, up to multiples. 
\end{rem}

If we go back to Algorithm \ref{alg:ghw}, assuming $m$ disjoint information sets, we may approximate the number of subspaces to be enumerated by
\begin{equation}\label{eq:enumeracionBW}
m\sum_{w=r}^{\lceil d/m-1 \rceil} e_w^r,
\end{equation}
where this is obtained as the number of subspaces that we need to enumerate to get the lower bound $m(w+1)$ to be greater than or equal to $d$. If we now approximate $e_w^r$ by $\binom{k}{w} { w\brack r}_q$, it is clear that the number of subspaces to be enumerated with this algorithm is, in general, much lower than that of the naive algorithm in which we enumerate ${ k\brack r}_q=\sum_{w=r}^k e_w^r$ subspaces. 

In any case, this computation of $e_w^r$ does not give a way to construct $E_w^r$ in an efficient manner, since the algorithm derived from this enumeration would just compute all subspaces of dimension $r$ with length $w$, and then remove those without full support. We show now one way to construct these subspaces efficiently.

We will first construct all the subspaces $E\subset\fq^k$ with $\dim E=r$ and $\supp(E)=\{1,\dots,w\}$. We can recover all the subspaces $E\subset\fq^k$ with $\supp(E)=\{j_1,\dots,j_w\}$ by mapping the coordinates $\{1,\dots,w\}$ to $\{j_1,\dots,j_w\}$. This can also be seen as adding columns of zeroes in appropriate positions. Thus, once we construct all the subspaces $E\subset\fq^k$ with $\dim E=r$ and $\supp(E)=\{1,\dots,w\}$, we can derive $E_w^r$.

Note that we can associate an unique matrix to each subspace $E\subset\fq^k$, namely the reduced row echelon form of any matrix whose rows are a basis for $E$. We denote this matrix by $R(E)$. Thus, the problem we want to solve is equivalent to enumerating all the reduced row echelon forms with $r$ rows and $k$ columns, with rank $r$, and such that all the first $w$ columns are nonzero and the last $k-w$ are zero columns. 

Since the last $k-w$ columns are zero, without loss of generality we may think about matrices with $w$ columns instead (or, in other words, we may assume $k=w$). Now we consider $i_1=1<i_2<\cdots i_r\leq w$. For each choice of $(i_1,\dots,i_r)$, we will construct the matrices $R(E)$ in reduced row echelon form whose pivots are in those positions. This determines $r$ of the columns of $R(E)$, and also the ``shape'' of the matrix:

$$ R(E)=
\begin{blockarray}{ccccccccccccccc}
  \begin{block}{(ccccccccccccccc)}
    1&*& \dots &*&0&*&\cdots&*&0&*&\dots &0&*&\dots&*\\
    0&0&\dots &0&1&*&\cdots &*&0&*&\dots &0&*&\dots&*\\
    0&0&\dots&0&0&0&\dots&0 &1&*&\dots &0 &*&\dots&*\\
    0&0 &\dots &0&0&0&\dots&0&0&0&\dots&0&*&\dots&*\\
    \vdots&\vdots&\ddots&\vdots&\vdots&\vdots &\ddots &\vdots &\vdots &\vdots &\ddots&\vdots &\vdots&\ddots&\vdots\\
    0&0&\dots &0&0&0&\dots&0&0&0&\dots &1 &*&\dots&* \\
  \end{block}
  \begin{block}{ccccccccccccccc}
    i_1 &&&&i_2&&&&i_3&&&i_r&&&w  \\
  \end{block}
\end{blockarray}\;.
$$
Thus, we still have freedom to choose $w-r$ columns. For each $\ell \in \{1,\dots,w\}\setminus \{i_1,\dots,i_r\}$, if $i_z< \ell < i_{z+1}$ (we define $i_{r+1}:=w+1$), we have to choose a column  $v_\ell\in \fq^r$ with $1\leq \wt(v_\ell)\leq z$. Indeed, $1\leq\wt(v_\ell)$ since all the columns have to be nonzero to have $\supp(E)=\{1,\dots,w\}$, and $\wt(v_\ell)\leq z$ because of the shape of $R(E)$. Each choice of the $w-r$ columns $v_\ell$ gives a different matrix $R(E)$, which corresponds to a different subspace $E$. Similarly, the matrices constructed with different sets of pivots correspond to different subspaces. Thus, in this way we construct every subspace $E$ with $\dim E=r$ and $\supp(E)=\{1,\dots,w\}$, without repeating. The corresponding pseudo-code is given in Algorithm \ref{alg:RE}. As mentioned before, from this we can recover $E^w_r$ by adding columns of zeroes in suitable positions. 

To consider all possible columns $v_\ell$, we need to construct all the vectors $v_\ell\in \fq^r$ with $1\leq \wt(v_\ell)\leq z$. It is straightforward to obtain these vectors by constructing all $v_\ell\in \fq^r$ with $\wt(v_\ell)=y$, for any $1\leq y \leq z$. This can be done by first selecting the position of the $y$ nonzero coordinates (there are $\binom{r}{y}$ possibilities), and then selecting an element of $\fq^*=\fq\setminus \{0\}$ for each position ($(q-1)^y$ possibilities). We now have all the tools we need to implement Algorithms \ref{alg:ghw} and \ref{alg:hierarchy} (and their relative version).

\begin{algorithm}
\caption{Construction of $R(E)$, for all $E\subset\fq^k$ with $\dim E=r$ and $\supp(E)=\{1,\dots,w\}$}\label{alg:RE}
\begin{algorithmic}[1] %[1] for numbering of lines
\Input $q, k, w,r$.
\Output List $L$ of $R(E)$ for all $E\subset \fq^k$ with $\dim E=r$ and $\supp(E)=\{1,\dots,w\}$.
\State $L=[\;]$
\For {$1=i_1<i_2<\dots <i_r\leq w$}
    \For {$\ell \in \{1,\dots,w\}\setminus \{i_1,\dots,i_r\}$}
        \If {$i_z<\ell <i_{z+1}$}
            \State $\text{cols}[\ell]=\{v\in \fq^r ,\; 1\leq \wt(v)\leq z \}$
        \EndIf
    \EndFor
    \State Define $\text{cart}$ as the Cartesian product of the sets in cols. Assume the entries of $\text{cart}$ are still indexed by $\ell \in \{1,\dots,w\}\setminus \{i_1,\dots,i_r\}$.
    \For{$u\in \text{cart}$}
        \For{$1\leq \ell \leq w$}
            \If{$\ell =i_z$ for some $1\leq z\leq r$}
                \State $M[\ell]=(0,\dots,0,1,0,\dots,0)$, where the $1$ is in position $z$.
            \Else
                \State $M[\ell]=u[\ell]$
            \EndIf
        \EndFor
        \State Append the matrix whose columns are given by $M$ to $L$.
    \EndFor
\EndFor
\State \Return $L$
\end{algorithmic}
\end{algorithm}

\section{Implementation}\label{s:implementation}
For the implementation, we have chosen Sage \cite{sagemath}. Sage is a free and open-source mathematics software with a Python-based language. Since it is free, anyone will be able to use this implementation, and the fact that Sage is mostly based on Python makes the code easier to understand. This is particularly important since this is the first implementation of these algorithms and it can serve as the starting point for possible future advances in computing GHWs and RGHWs. The fact that Sage includes functions and classes for all the objects we use, e.g., linear codes, also facilitates the understanding of the implementation. Additionally, this implementation aims to be as general as possible, working over any finite field and for any linear code. In particular, this means we need to use mathematical software with a good implementation of finite fields, which heavily restricts the options. The other two options to be considered in this regard would be Magma \cite{magma}, which is not open-source, and GAP \cite{gap}, which could be an alternative for Sage in this case, but it uses its own programming language. 

The package {\tt GHWs} includes several auxiliary functions and also some main functions. The auxiliary functions include the function {\tt information}, which computes the required information sets for Algorithms \ref{alg:ghw} and \ref{alg:hierarchy}, as well as the corresponding generator matrices and redundancies. This uses the procedure explained in Section \ref{s:BZalg}, and since this is also used in the usual Brouwer-Zimmermann algorithm, we shall not explain it further (see \cite{grasslBrouwerZimmermann}). In any case, the functions relying on {\tt information} have an optional argument to provide alternative information sets and generator matrices if the user wants to compute them in a different way. We also include some auxiliary functions to check if a code is cyclic and to compute its BCH bound, and some small functions required for the main algorithms. All the functions included have a description text explaining what they do and providing examples.

As main functions we have {\tt GHW}, {\tt hierarchy}, {\tt RGHW} and {\tt rhierarchy}, which compute the GHWs, hierarchy, RGHWs, and relative hierarchy, of a code (or a pair codes in the relative case). For the case of {\tt GHW} (the others are analogous), we first check if the code is cyclic to make use of the improvements related to cyclic codes (see Subsection \ref{ss:considerations} below), and we compute the information sets and generator matrices using {\tt information} otherwise. Then we proceed as in Algorithm \ref{alg:ghw}, but at the beginning of each iteration we make a prediction of the expected $w_0$ required to finish, meaning that $E_{w_0}^r$ would be the last set we expect to enumerate. We use this to remove unnecessary generator matrices if the lower bound at the end of this step will be much higher than the current upper bound (see Subsection \ref{ss:considerations} below). Then we compute $R(E)$ for all subspaces $E$ of $\fq^w$ with $\abs{\supp(E)}=w$ and $\dim E=r$ using {\tt subspaces(r, w, w, K)} (in this case, $K=\fq$). This follows the idea of Section \ref{s:enumeration}, and the function {\tt subspaces} computes all the required $R(E)$ by constructing the reduced row echelon forms as in Algorithm \ref{alg:RE}. We store all these matrices, since we use them to construct all the $R(E)$, for $E\in E_w^r$ by appropriately adding columns of zero's. In some cases, this may require a substantial RAM usage (for example, if $q$ is large). We provide an alternative implementation of these functions without storing these matrices ({\tt GHW_low_mem}, {\tt hierarchy_low_mem}, {\tt RGHW_low_mem} and {\tt rhierarchy_low_mem}). However, this is usually slower (see Section \ref{s:tests}) since we need to repeat some of the operations required for the construction of $E_w^r$ several times. After this, for each $E\in E_w^r$, we compute $\enc_{G_i}(E)$, for every generator matrix $G_i$ that contributes to the lower bound, and update the upper bound if the cardinality of the support is lower than the current upper bound. Finally, we compute the new lower bound at the end of the iteration. 

For the case of {\tt RGHW} and {\tt rhierarchy} (and their low memory version), we need to add an extra step: if $\abs{\supp(\enc_{G_i}(E))}$ is lower than the current upper bound, then we check if $\enc_{G_i}(E)\cap C_2=\{0\}$, and we only update the value of the upper bound if that is the case, which is clearly more efficient than checking the condition for every subspace that we consider. To check that $\enc_{G_i}(E)\cap C_2=\{0\}$, Sage has classes for vector spaces, and we can create the corresponding vector spaces and compute the dimension of the intersection. However, to avoid making these conversions, we note that checking $\dim(\enc_{G_i}(E)\cap  C_2)=0$ is equivalent to checking $\enc_{G_i}(E)\cap C_2=\{0\}$, and we use instead
$$
\dim(\enc_{G_i}(E)\cap  C_2)=r-\rk (H_2(R(E)G_i)^t),
$$
which can be derived from \cite[Prop. 2.2]{relativehull} ($H_2$ is a parity check matrix of $C_2$, that is, a generator matrix of $C_2^\perp)$. 

Finally, we have also included the function {\tt higher_spectrum} to compute the higher weight spectrum, and the function {\tt rhigher_spectrum} to compute the relative higher weight spectrum. For these functions, one necessarily has to enumerate all the subspaces of the code, so there is no improvement in the enumeration as in the Brouwer-Zimmermann-like algorithm, but we have included it since it can be directly computed using the function {\tt subspaces}. For the low memory version, we proceed similarly to what we did with {\tt GHW_low_mem}.

The functions {\tt GHW}, {\tt hierarchy}, {\tt RGHW}, {\tt rhierarchy}, {\tt higher_spectrum}, {\tt rhigher_spectrum}, and their low memory version, have a {\tt verbose} argument, which, if set to {\tt True}, it shows real time information about the computation. This is recommended for heavy computations, since it can give an idea of whether a computation will finish within a reasonable time or not. 

\subsection{Additional considerations}\label{ss:considerations}

Although our main focus is to compute the GHWs and RGHWs of a random linear code, we have included some improvements for cyclic codes, since they are straightforward to implement. One can check how those improvements work, for instance, in \cite{grasslBrouwerZimmermann} (it is analogous to the case of the minimum distance). In the same reference, one can also find that in some cases we might not use some of the additional generator matrices we are considering. For example, if $w+1-R_j\leq 0$ for some $j$ in Algorithm \ref{alg:ghw}, that matrix does not improve the lower bound and we may skip it. Moreover, we can predict what the value of the lower bound is going to be at the end of the iteration, and if the value surpasses the current upper bound by a sufficiently large margin, we may also skip some matrices while still ensuring we finish the algorithm in this iteration (since the lower bound can still be greater than or equal to the current upper bound). These improvements for the last iteration of the algorithm have a great impact, since it is usually the most computationally expensive step. For example, see \cite[Lem. 2]{hernandoFastBZRandomLinearCode}, in which it is shown that in some cases this iteration is more expensive than all the previous ones, for $r=1$ and $q=2$.

Due to Wei's duality from Theorem \ref{t:ghwdual}, computing the weight hierarchy of a code is equivalent to computing the weight hierarchy of its dual, and we have the function {\tt wei_duality} to compute this. In many cases it will be more efficient to compute the weight hierarchy of the dual of a code and then use Wei's duality than computing it directly. Technically, we do not know in advance exactly how many computations will be needed. However, since the size of $e_w^r$ is essentially given by $\binom{n}{w}{ w\brack r}_q $, for $1\leq r \leq k$ and $r\leq w \leq k$, a good strategy would be to compute the weight hierarchy directly if $k\leq n/2$, and compute the weight hierarchy of the dual if $k>n/2$. 

\section{Tests}\label{s:tests}

For the results presented in this section, we have used Arch Linux on a computer with Windows 10 using Windows Subsystem for Linux (WSL). The virtual machine has access to 7.72GB of RAM and an Intel Core i7-6700 processor (3.4GHz, with Turbo Boost disabled). With respect to Sage, we have used the 10.6 version. We also had access to a server with an AMD EPYC 7F52 processor (3.5GHz, up to 3.9GHz with boost), with Ubuntu 20.04.6 LTS and SageMath 10.6, obtaining similar results. We provide several test files to replicate our results. For example, for the test file {\tt test_GHWs.sage}, the corresponding test is ran by writing {\tt sage test_GHWs.sage}. These test files assume that the folder structure follows the one given in \cite{githubGHWs}. Otherwise, the line {\tt load(path2)} in each test file has to be changed to specify the path of {\tt GHWs.py} (these files can be modified with a standard text editor).

\subsection{Correctness}

Since this is a new algorithm, the first thing to test is the correctness of the results. For this, we may consider families of codes with known GHWs and RGHWs, and check that our algorithms return the correct values. By Remark \ref{r:GHWMDS}, the weight hierarchy of MDS codes is known, and their higher weight spectra has been studied in \cite{doughertyHigherWeights}. The most well-known class of MDS codes is given by Reed-Solomon (RS) codes. For a given $q$ and a given dimension $k$, the corresponding RS code is denoted by $\RS_q(k)$ and has parameters $[q,k,q-k+1]$. These codes require working over large finite fields due to the limitation on the length, requiring a substantial number of operations for the algorithms to finish. 

Another family with known weight hierarchy are Reed-Muller (RM) codes \cite{pellikaanGHWRM,kasamiRM}, and their RGHWs have also been studied in \cite{geilRGHWReedMuller}. For a given integer $m>0$ and a degree $0\leq \nu <q$, the corresponding RM code is denoted by $\RM_q(\nu,m)$, and has parameters $\left[q^m,\binom{d+m}{m},(q-d)q^{m-1}\right]$ (for $\nu\geq q$, their parameters are also known but are more complicated). Both RS and binary RM codes are directly implemented in Sage, as well as $q$-ary RM codes for $\nu <q$. The advantage of RM codes in this case is that we can have longer codes over a smaller finite field. Similarly, cyclic codes are implemented in Sage, some of them have known GHWs \cite{janwaGHWcyclic}, and they can provide long codes over small finite field sizes. In the test file {\tt test_GHWs.sage}, we have included several tests to check that our implementation correctly gives the GHWs of RS, RM and cyclic codes. If no error is returned, the test has been successfully passed, and this can be used to also check that all the functions are working properly (a similar test can be ran with the test file {\tt test_GHWs_low_mem.sage} for the low memory functions). 

In particular, these tests check the values returned by our main functions compared with the known GHWs of some binary Reed-Muller codes, the RGHWs of some $q$-ary Reed-Muller codes (with $q=5$), the GHWs of some binary cyclic codes, and the GHWs, Wei's duality from Theorem \ref{t:ghwdual}, higher spectra and relative higher spectra of some Reed-Solomon codes. The test {\tt test_GHWs.sage} takes 226.27s on average on our machine, and the test {\tt test_GHWs_low_mem.sage} takes 406.06s. This already shows that the low memory implementation can be slower, and we will explore this later.

\subsection{Performance analysis}

In terms of performance, since there are no alternatives currently, we cannot compare with other implementations. The only test that we can make in this direction is to compare with the naive approach of enumerating all the subspaces. The most naive approach would just consider all $r\times k$ matrices over $\fq^k$ with rank $r$ and then compute the minimum support of the resulting matrices after multiplying by $G$, a generator matrix of $C$. Instead of this, we can use our function {\tt subspaces}, which will not repeat subspaces (as the previous method would do, since there are several matrices whose row span is the same). However, compared to our Brouwer-Zimmermann-like approach, this naive algorithm may require to compute many more subspaces (recall (\ref{eq:enumeracionBW})). This becomes problematic in terms of RAM for medium to large finite field sizes. Thus, in the test file {\tt test_change_r} we have included the function {\tt naive_GHW} (using {\tt subspaces} and requiring more RAM), and a version in the spirit of {\tt GHW_low_mem}, called {\tt naive_GHW_low_mem}, which requires less RAM. For our tests, we shall only use {\tt naive_GHW_low_mem}, and, to make a fair comparison, we compare with {\tt GHW_low_mem} (in the test files {\tt test_change_k.sage} and {\tt test_change_q.sage} we only include {\tt naive_GHW_low_mem}). Also, to avoid having to handle lists of generator matrices of random codes, we will use families of codes that are implemented in Sage, as before. Experiments with random codes show similar performance for our functions.

\begin{figure}
\caption{Semi-logarithmic plots of Tables \ref{t:changek} and \ref{t:changeq}.}\label{f:changekq}
    \centering
    \resizebox{0.49\linewidth}{!}{\includegraphics{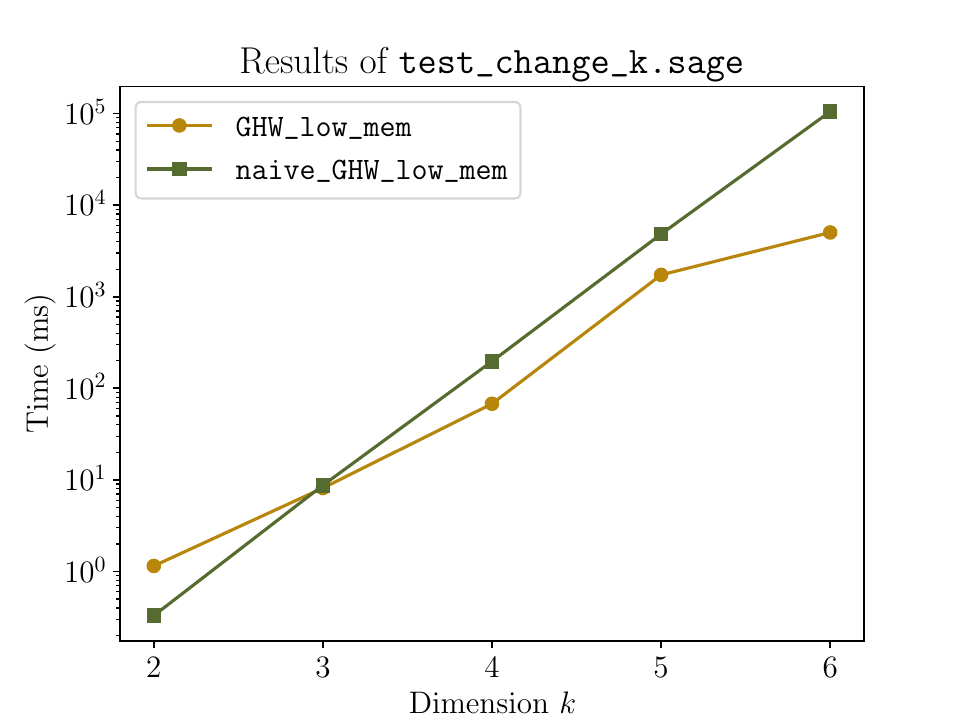}}
    \resizebox{0.49\linewidth}{!}{\includegraphics{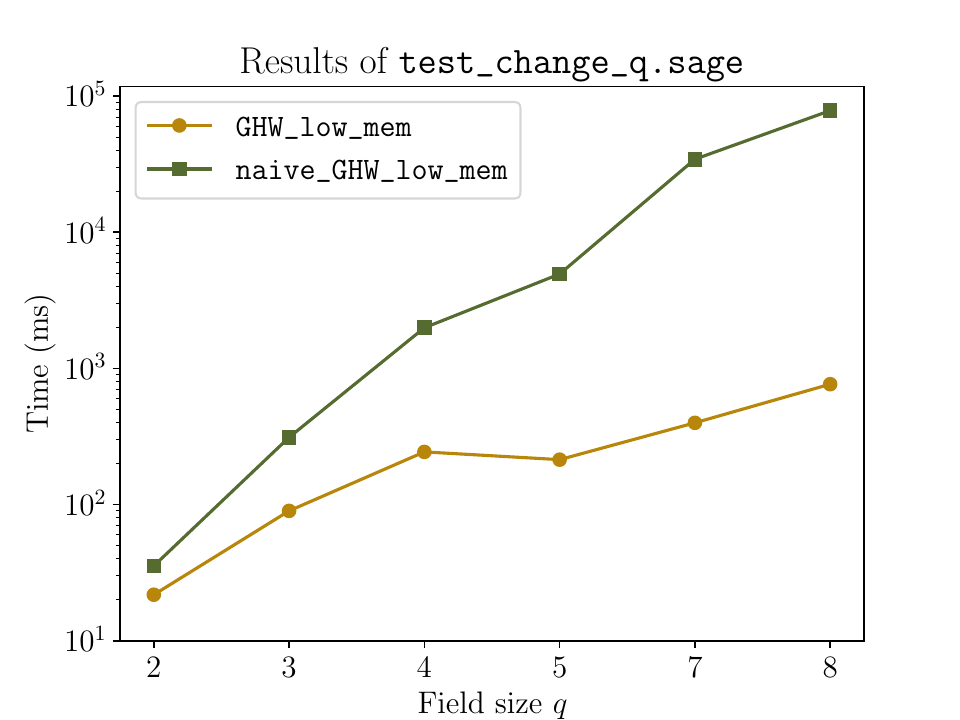}}
\end{figure}

First, with {\tt test_change_r.sage} we have compared the times required for computing the GHWs of RM codes with $q=5,7$, and $r=2,3,4,5$, using {\tt GHW_low_mem} and {\tt naive_GHW_low_mem}. The results are given in Tables \ref{t:changer5} and \ref{t:changer7}. We see that the difference in performance is substantial (in some cases there are two orders of magnitude of difference). This will also be the case in many of the subsequent tests, which is why we have chosen tables instead of plots to present most of our results. Since both implementations are in Sage, the difference in time represents the advantage gained by using our Brouwer-Zimmermann-like algorithm with respect to just enumerating all subspaces. All these codes have the same dimension $k=6$, and for a fixed $q=5$ or $q=7$, we only change $r$. With {\tt test_change_k.sage}, we study the dependence on $k$ by fixing $r=2$ and $q=5$, and computing the second GHW of certain subcodes of the code $\RM_5(3,2)$ with different dimensions $k$. The results are presented in Table \ref{t:changek}. We see that, although the time required for both functions increases rapidly, the increase on time required for {\tt naive_GHW_low_mem} to finish is much larger than that of {\tt GHW_low_mem}. We also note that, for very low values of $k$, the naive approach can be faster since in that case using several generator matrices may not be beneficial. But these cases are not our main focus, as both functions will complete in a short amount of time. Finally, with {\tt test_change_q.sage}, we consider the dual Hamming code with redundancy $5$. This code is a binary code, which means it has a generator matrix given by zeroes and ones. We can consider this matrix as the generator matrix of a code over any finite field, and all the resulting codes will have the same length and dimension. We have computed the second GHW for these codes, for $q=2,3,4,5,7,8$, and the results are shown in Table \ref{t:changeq}. Clearly, the size of the field seems to impact the performance of the naive approach much more than the performance of {\tt GHW}. We have also represented the results of Tables \ref{t:changek} and \ref{t:changeq} in Figure \ref{f:changekq}. Note that we are using a logarithmic scale for the $y$ axis due to the exponential difference in time (and exponential growth of the time required to finish).

\begin{table}[ht]
% title of Table
\caption{Time in seconds to compute $d_r(\RM_5(2,2))$, $2\leq r \leq 5$.}\label{t:changer5}
\centering
%\begin{center}
\begin{tabular}{c|cccc}
 \hline % \hline
  % after \\: \hline or \cline{col1-col2} \cline{col3-col4} ...
Function &$d_2(\RM_5(2,2))$&$d_3(\RM_5(2,2))$&$d_4(\RM_5(2,2))$&$d_5(\RM_5(2,2))$ \\
  \hline 
{\tt GHW_low_mem}& 3.840 & 1.165&2.564& 0.01116 \\
{\tt naive_GHW_low_mem}& 89.840 & 577.613 &139.910 &1.55701
\\
\hline
\end{tabular}
\end{table}

\begin{table}[ht]
\caption{Time in seconds to compute $d_r(\RM_7(2,2))$, $2\leq r \leq 5$.}\label{t:changer7}
\begin{tabular}{c|cccc}
 \hline % \hline
  % after \\: \hline or \cline{col1-col2} \cline{col3-col4} ...
Function &$d_2(\RM_7(2,2))$&$d_3(\RM_7(2,2))$&$d_4(\RM_7(2,2))$&$d_5(\RM_7(2,2))$ \\
  \hline 
{\tt GHW_low_mem}& 517.4 & 714.6 & 20.08 & 0.03101  \\
{\tt naive_GHW_low_mem} & 1,342.5 & 12,519.0 & 2,287.91 & 11.90103 \\
\hline
\end{tabular}
\end{table}

\begin{table}[ht]
% title of Table
\caption{Time in milliseconds to compute $d_2(C)$, where $C\subset \RM_5(3,2)$ with $\dim C=k$ (the definition of $C$ can be found in {\tt test_change_k.sage}).}\label{t:changek}
\centering
%\begin{center}
\begin{tabular}{c|ccccc}
 \hline % \hline
  % after \\: \hline or \cline{col1-col2} \cline{col3-col4} ...
Function &$k=2$&$k=3$&$k=4$&$k=5$ &$k=6$ \\
  \hline 
{\tt GHW_low_mem}& 1.15 & 8.15&67.81& 1,731.14 & 5,031.54\\
{\tt naive_GHW_low_mem}& 0.33 & 8.73 & 197.21 & 4,831.71& 105,041.21
\\
\hline
\end{tabular}
\end{table}

\begin{table}[ht]
% title of Table
\caption{Time in milliseconds to compute $d_2(C)$, where $C$ is the Hamming code with redundancy 5 over $\F_2$, seen as a code over $\fq$, for $q=2,3,4,5,7,8$.}\label{t:changeq}
\centering
%\begin{center}
\begin{tabular}{c|cccccc}
 \hline % \hline
  % after \\: \hline or \cline{col1-col2} \cline{col3-col4} ...
Function &$q=2$&$q=3$&$q=4$&$q=5$ &$q=7$ & $q=8$ \\
  \hline 
{\tt GHW_low_mem}& 21.8 & 90.1 & 244.5& 214.3& 399.6& 766.7\\
{\tt naive_GHW_low_mem}& 35.5& 312.7 & 1,991.3& 4,943.4& 34,423.0& 78,320.1
\\
\hline
\end{tabular}
\end{table}

Overall, we see that the our algorithm performs much better than the  naive approach. Note that for the naive approach we are also using part of our code to construct all subspaces without repeating (Section \ref{s:enumeration}). If we do not use these ideas, and we just consider all $r\times k$ matrices over $\fq^k$ with rank $r$ (what we called before ``the most naive approach''), since every subspace of dimension $r$ has $(q^r-1)(q^r-q)\cdots (q^r-q^{r-1})$ different bases, we would be multiplying the number of operations by this number (we would consider different matrices with the same row span), making the computation of the GHWs unfeasible in most nontrivial cases. 

\begin{figure}[ht]
    \centering
    \caption{Time in seconds to compute the GHWs of some RM codes using {\tt GHW} and {\tt GHW_low_mem}.}\label{f:normal_low_mem}
    \resizebox{0.75\linewidth}{!}{\input{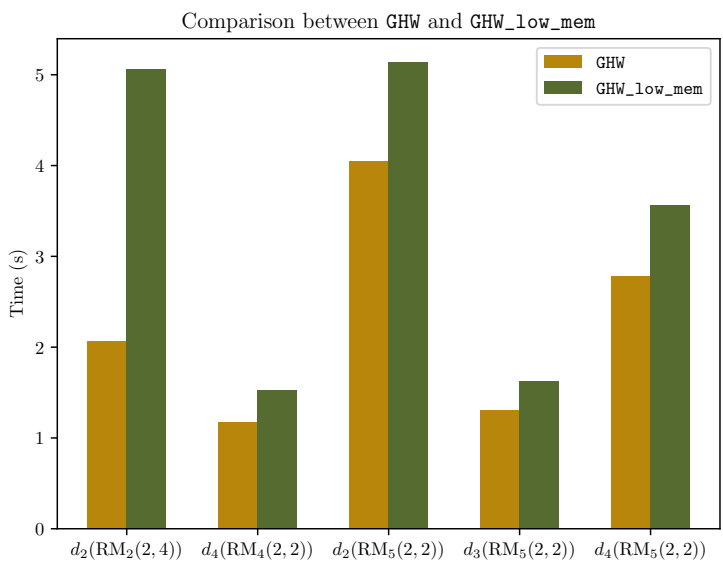}}
    \vspace{-0.75cm}
\end{figure}

\begin{figure}[ht]
    \centering
    \caption{Peak RAM usage of {\tt GHW} for several RM codes.}\label{f:ram}
    \resizebox{0.9\linewidth}{!}{\input{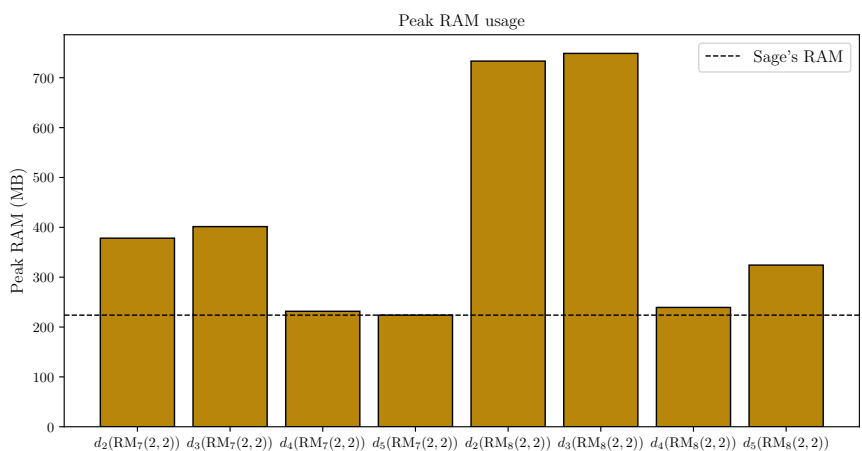}}
    \vspace{-0.75cm}
\end{figure}

We study now the performance of {\tt GHW} with respect to {\tt GHW_low_mem} (similar results are obtained when comparing {\tt hierarchy} with {\tt hierarchy_low_mem}, etc). We have already explained theoretically that, in general, {\tt GHW} should be faster, at the cost of requiring more RAM, and we have seen with {\tt test_GHWs.sage} and {\tt test_GHWs_low_mem.sage} that this is indeed the case. With {\tt test_normal_low_mem.sage}, we check the speed of both functions for some RM codes of degree $d$, for different field sizes and values of $r$. The results are shown in Figure \ref{f:normal_low_mem}. The difference in performance varies, but in general {\tt GHW} is faster than {\tt GHW_low_mem}. With respect to RAM, {\tt GHW_low_mem} uses a negligible amount of memory besides the memory required to load Sage, which on our machine was about 2.9\% of the total RAM (223.88MB). In Figure \ref{f:ram} we show the peak RAM usage of {\tt GHW} while computing the GHWs of several RM codes (for low values of $q$, the memory usage was negligible). For $r=2,3$, which are the values for which the list of subspaces required is larger, we see that the RAM requirement can be noticeable when we increase the finite field size $q$. In most of our experiments, time became an issue before RAM did, in the sense that the cases with a large RAM usage were not expected to finish in a reasonable amount of time, but this will depend on the relation between available RAM and processing power of each machine.

We can also make a comparison between {\tt GHW} and {\tt hierarchy} for computing the weight hierarchy of a linear code (similarly for the relative version). The only difference between using {\tt GHW(C,r)}, for $1\leq r \leq k$, and {\tt hierarchy(C)}, is that {\tt hierarchy} can use the value $d_r(C)+1$ as a lower bound for $d_{r+1}(C)$. Depending on the code, this can be a negligible improvement or a very important one. The most extreme case is given by MDS codes, since {\tt hierarchy} will end after computing $d_1(C)$ because at that point the generalized Singleton bound from Corollary \ref{c:singletongeneralizada} will meet the lower bound for every $2\leq r \leq k$. We show this in Table \ref{t:hierarchyRS}, where we compute the weight hierarchy of $\RS_{13}(k)$ for different values of $k$. For other families of codes, the improvement will not be as large, as we see in Table \ref{t:hierarchyRM}, where we have used RM codes instead. These tests can be found in {\tt test_hierarchy_ghw.sage}.

\begin{table}[ht]
% title of Table
\caption{Time in milliseconds to compute the weight hierarchy of $\RS_{13}(k)$, for $1\leq k \leq 12$.}\label{t:hierarchyRS}
\centering
%\begin{center}
\begin{tabular}{c|cccccc}
 \hline % \hline
  % after \\: \hline or \cline{col1-col2} \cline{col3-col4} ...
Function &$k=1$&$k=2$&$k=3$&$k=4$&$k=5$ &$k=6$ \\
  \hline 
{\tt hierarchy}& 0.557 & 0.846&6.60& 11.21& 173.2& 323.6\\
{\tt GHW}& 0.559 & 1.654 & 9.77& 138.8& 452.3 & 6751.9 \\
\hline
&$k=7$& $k=8$&$k=9$&$k=10$&$k=11$&$k=12$  \\
\hline
{\tt hierarchy}& 547.7 & 891.2 & 48.5 & 53.8 & 3.54 & 3.61\\
{\tt GHW}&  14,803.4 & 29,319.4 & 47,821.7 & 3,061.1 & 463.5& 1,001.0\\
\hline
\end{tabular}
\end{table}

\begin{table}[ht]
% title of Table
\caption{Time in milliseconds to compute the weight hierarchy of some RM codes.}\label{t:hierarchyRM}
\centering
%\begin{center}
\begin{tabular}{c|cccc}
 \hline % \hline
  % after \\: \hline or \cline{col1-col2} \cline{col3-col4} ...
Function & $\RM_2(2,3)$ & $\RM_3(2,2)$ & $\RM_4(2,2)$ & $\RM_5(2,2)$ \\
  \hline 
{\tt hierarchy}& 1.99& 37.77 & 2,770.1 & 5,634.0 \\
{\tt GHW}& 22.31 & 153.41 & 3,443.3 & 6,933.2 \\
\hline
\end{tabular}
\end{table}

Since our functions are completely general, we do not expect them to be competitive for the case of the minimum distance ($r=1$) where additional improvements can be considered \cite{hernandoFastBZRandomLinearCode,hernandoBZQuantum,hernandoParallelBZ}. Moreover, if one restricts the field size to $q=2,3$, then it is possible to use lower level programming languages, which will result in a better performance. Notwithstanding the foregoing, in our experiments {\tt GHW(C,1)} has performed better than the Sage function {\tt C.minimum_distance()} for higher values of $k$, although it is slower for lower values of $k$. In Sage, one can also use {\tt C.minimum_distance(algorithm='guava')} to use the algorithm from GUAVA \cite{guava}. GUAVA includes an implementation of the Brouwer-Zimmermann algorithm for $q=2,3$ in C, which should be faster than our implementation for $r=1$. Nevertheless, it is noteworthy that in some cases our implementation can be faster. For example, if one considers the BCH code obtained by writing {\tt C = codes.BCHCode(GF(2), 2^7-1, 27)} in Sage (BCH code over $\F_2$ of length 127 and designed minimum distance 27), {\tt GHW(C,1)} takes 305ms on average to finish on our machine, while {\tt C.minimum_distance(algorithm='guava')} takes 19.4s on average. We will not analyze the case of $r=1$ further, since our main objective is to implement the algorithm in the most general form for any $r$, but the fact that our implementation outperforms specialized implementations for the case $r=1$ in some cases is an indication of its efficiency. 

\section{Conclusion}\label{s:conclusion}
In this work we have introduced a generalization of the Brouwer-Zimmermann algorithm for computing the GHWs of a linear code, which required studying how to construct all subspaces with a given support. The resulting algorithm has been implemented in Sage, and we have shown that this implementation is much more efficient than computing the GHWs using the definition. 

In the future, we will study improvements for particular cases. The case of $q=2$ is particularly interesting since we may use a lower level programming language, like C, and the construction of all the subspaces may be optimized due to working with such a small finite field size. We may also consider particular values of $r$, such as $r=2$, which, as we have mentioned earlier, appears in some additional applications, e.g., determining the error-correction capability of quantum codes derived via Steane enlargement \cite{hamadaSteaneEnlargement}.

% \bibliographystyle{abbrv}
% \bibliography{BIBR} 

\end{document}